\documentclass{article}

\usepackage[final]{neurips_2021_ml4ps}

\usepackage[utf8]{inputenc} 
\usepackage[T1]{fontenc}    
\usepackage{url}            
\usepackage{booktabs}       
\usepackage{amsfonts}       
\usepackage{nicefrac}       
\usepackage{microtype}      
\usepackage{multirow}
\usepackage{ctable}
\usepackage{multicol}
\usepackage{bm}

\definecolor{darkpastelgreen}{rgb}{0.01, 0.75, 0.24}

\title{Robustness of deep learning algorithms in astronomy - galaxy morphology studies}

\author{
\hspace*{-2mm}
A. \'Ciprijanovi\'c\\
\hspace*{-2mm}
\texttt{aleksand@fnal.gov}\\
\hspace*{-2mm}
Fermi National Accelerator Laboratory\And

\hspace*{4mm}
D. Kafkes\\
\hspace*{4mm}
\texttt{dkafkes@fnal.gov}\\
\hspace*{4mm}
Fermi National Accelerator Laboratory\And

\hspace*{-7mm}
G. N. Perdue\\
\hspace*{-7mm}
\texttt{perdue@fnal.gov}\\
\hspace*{-7mm}
Fermi National Accelerator Laboratory\And

\hspace*{-7mm}
K. Pedro\\
\hspace*{-7mm}
\texttt{pedrok@fnal.gov}\\
\hspace*{-7mm}
Fermi National Accelerator Laboratory\And

\hspace*{9mm}
G. Snyder\\
\hspace*{9mm}
\texttt{gsnyder@stsci.edu}\\
\hspace*{9mm}
Space Telescope Science Institute\And

\hspace*{9mm}
F. J. S\'{a}nchez\\
\hspace*{9mm}
\texttt{jsanch87@fnal.gov}\\
\hspace*{9mm}
Fermi National Accelerator Laboratory\And

\hspace*{2mm}
S. Madireddy\\
\hspace*{2mm}
\texttt{smadireddy@anl.gov}\\
\hspace*{2mm}
Argonne National Laboratory\And

\hspace*{2mm}
S. M. Wild\\
\hspace*{2mm}
\texttt{wild@anl.gov}\\
\hspace*{2mm}
Argonne National Laboratory\And

\hspace*{9mm}
B. Nord\\
\hspace*{9mm}
\texttt{nord@fnal.gov}\\
\hspace*{9mm}
Fermi National Accelerator Laboratory,\\
\hspace*{9mm}
Kavli Institute for Cosmological Physics,\\ 
\hspace*{9mm}
University of Chicago \\
\hspace*{9mm}
Department of Astronomy and Astrophysics,\\
\hspace*{9mm}
University of Chicago \\

}

\begin{document}

\maketitle

\begin{abstract}
Deep learning models are being increasingly adopted in wide array of scientific domains, especially to handle high-dimensionality and volume of the scientific data. However, these models tend to be brittle due to their complexity and overparametrization, especially to the inadvertent adversarial perturbations that can appear due to common image processing such as compression or blurring that are often seen with real scientific data. It is crucial to understand this brittleness and develop models robust to these adversarial perturbations. To this end, we study the effect of observational noise from the exposure time, as well as the worst case scenario of a one-pixel attack as a proxy for compression or telescope errors on performance of ResNet18 trained to distinguish between galaxies of different morphologies in LSST mock data. We also explore how domain adaptation techniques can help improve model robustness in case of this type of naturally occurring attacks and help scientists build more trustworthy and stable models. 
\end{abstract}

\section{Introduction}
\label{sec:intro}

While the ability of neural networks to perform well on a broad range of tasks is remarkable, there has been a recent push to evaluate instances when neural networks fail. The study of adversarial examples seeks to explore this brittleness through the context of specifically-crafted inputs designed to confuse the network, sometimes leading to catastrophic results. Some attacks rely on the ability to access information about the networks themselves, such as the underlying architecture, trained weights, and gradients [25, 11]. On the other hand black box attacks do not require any information about the trained network, but they can still be very successful in fooling the model [19, 1]. It is also known that even image compression, blurring, or even the addition of simple Gaussian noise can also wreak havoc on a network's performance [10, 6, 7, 8]. We posit that this sort of "attack" can arise quite easily in the sciences. For example, when simulating data in astronomy we can never perfectly model observational effects, even the most important ones like observational noise or blurring by the point spread function (PSF) of the telescope. Understanding these effects and ultimately finding solutions that will lead to more robust models is part of both adversarial example and domain adaptation research [4, 27, 29].

Here we explore the sorts of adversarial examples that might arise out of complex scientific data pipelines, such as the Vera C. Rubin Observatory’s Legacy Survey of Space and Time (LSST) [15], within the context of classifying galaxy morphology-- distinguishing between spiral, elliptical and merging galaxies. We produce LSST mock data using Illustris TNG100 [18] images, and study the effects of observational noise from the exposure time-- one year (Y1) vs. ten years (Y10) of observations. Additionally, we investigate the worst case scenario of a one-pixel attack as a proxy for compression or telescope errors. We study how domain adaptation methods, first introduced in the astronomical context in [2, 3], can be used to increase network robustness to these naturally accruing "adversarial attacks".

\section{Data}
\label{sec: Data}

We extract three-filter ($g,r,i$) galaxy images from snapshots 95 ($z=0.05$) and 99 ($z=0$) from the IllustrisTNG100 simulation [18]. We convert all data to effective redshift of $z=0.05$ and use the GalSim [21] package to make Y10 and Y1 LSST mocks by adding observational noise drawn from a Poisson distribution (made from a single image with varying exposure time) and PSF blurring (both atmospheric and optical). We aim to train a model to distinguish between spiral, elliptical and merging galaxies. To produce labels, we use IllustrisTNG morphology catalogs produced by [20] and follow [17] by using $G-M_{20}$ statistics to roughly distinguish between the three types of galaxy morphologies. Since non-merging galaxies are much more numerous in the final simulation snapshots, we also augment extracted mergers by flipping up/down and rotating images by $90^{\circ}$ and $180^{\circ}$, to get a more balanced dataset. This leads to our final galaxy sample with $14312:8151:12710$ images of spirals, ellipticals and mergers, respectively. We divide these images into training validation and test datasets as $70\%:10\%:20\%$. For example images, see Figure 1.

\section{Methods}
\label{sec: Methods}

We train ResNet18 network, the smallest standard "off-the-shelf" residual neural network [14], often used for different astronomy applications. After the convolutional layers, which extract interesting features, we have a $256$-dimensional bottleneck layer, which represents the latent space of the network we want to study. After the bottleneck layer we only have the output layer with three neurons representing the three galaxy classes. We train with Adam optimizer [16], use early stopping and fix random seed when training. Training was performed on Google Cloud's NVIDIA Tesla V100 GPU.

\begin{figure}
	\includegraphics[width=\columnwidth]{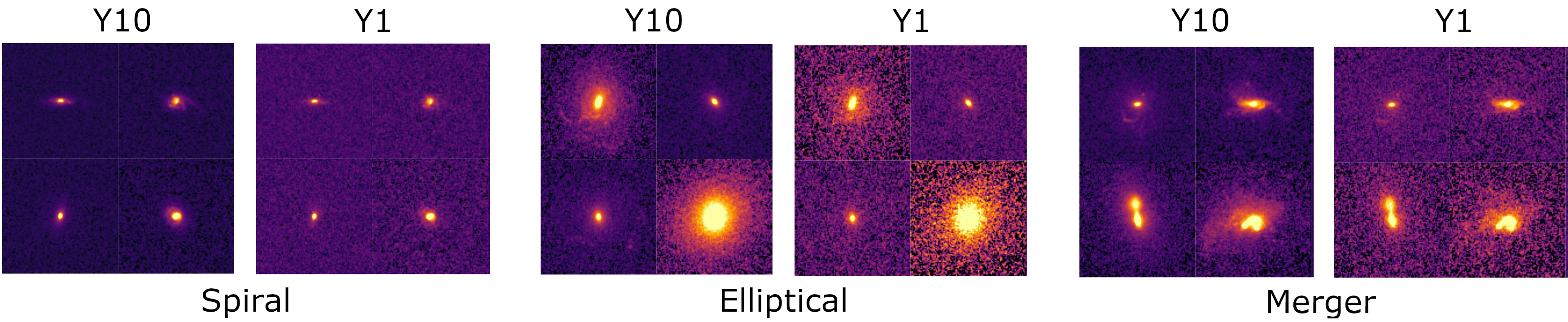}\\
    \caption{Examples of images in our dataset. The left two images show the same Y10 and Y1 examples of spiral galaxies, middle two images show Y10 and Y1 elliptical galaxies, while right two images show Y10 and Y1 examples of merging galaxies.}
\end{figure}

\textbf{Observational noise and exposure time} - as stated above, we create two versions of our data. For our main dataset, we add noise to mimic LSST observations after the full ten year observation period (Y10). To show how noise can affect deep learning performance, even in the case where the noise model is the same, and the only difference is just its signal to noise level, we also create mock catalogs of one year of LSST observations (Y1). 

\textbf{One-Pixel Attack} - we choose to explore adversarial perturbations due to one-pixel attack, since it is not unreasonable to expect that algorithms involved in compressing and decompressing large amounts of data could be expected to accidentally flip one pixel. Beyond compression a single pixel could easily be changed due to some problem with the charge-coupled device (CCD) detector or its readout, by effects of cosmic-rays, or through some other transient phenomena. The concept of a one-pixel attack was first introduced in [24]. This attack is a type of a "black-box" attack, which does not depend on the knowledge of the weights of the trained model. In [24], authors show that even widely used benchmarking datasets like CIFAR-10 and ImageNet~\footnote{https://www.cs.toronto.edu/\%7Ekriz/cifar.html; https://image-net.org/index.php} can be flipped to a wrong class by attacking just one pixel of the image. We can expect that noisy astronomical images will be susceptible to the same kind of problem as well. If we represent the original image as a tensor $\bm{x}$, and the probability of that image being classified into a particular class as $f(\bm{x})$, the one-pixel attack is optimized to find the additive perturbation vector $e(\bm{x})$, such that it maximizes the probability $f_\mathrm{adv}(\bm{x}+e(\bm{x}))$, of the image being misclassified as a different incorrect class. Finding the one pixel that leads to the desired perturbation is done by using a population based algorithm called differential evolution [24, 5].

\subsection{Domain Adaptation}
\label{sec: domain_adapt}
Big  part  of the adversarial attack research  is related to finding solutions (before or after the attack) and increasing model robustness [30]. In this work, we hypothesize that that domain adaptation techniques, which are used when the network trained on images from one domain needs to perform well in a new (often unlabeled) domain [4, 27, 29, 2], can also be very useful for increasing robustness to these naturally occurring attacks. We employ domain adaptation technique called Maximum Mean Discrepancy (MMD) [13, 22, 12]. MMD is a statistical technique that finds a non-parametric distance between two probability distributions (in this case latent space data distributions of our Y10 and Y1 images). By training a network such that this distance is minimized, MMD aligns the two data distributions in the latent space of the network, allowing the network to find domain-invariant features. To do this, instead of just minimizing standard classification cross-entropy loss ${\cal L}_\mathrm{CL}$, during training we minimize ${\cal L}_\mathrm{TOT} = {\cal L}_\mathrm{CL} + \lambda {\cal L}_\mathrm{MMD}$, where we use domain adaptation weight $\lambda=0.05$. In our case we will treat Y10 images as the labeled domain and noisy Y1 images as an unlabeled domain (note that domain adaptation methods will work even if domains are reversed - a realistic situation where we first observe Y1 data and then get new unlabeled Y10 data). Aligning Y10 and Y1 data will increase the noise robustness and should in principle also help with one-pixel attacks. We note that although we adopt a vanilla ResNet18 to carryout our domain adaption for robustness experiments, other adversarial robustness approaches such as those based on architecture improvement and data augmentation can be used alongside domain adaptation. This is a future direction we will pursue.

\section{Results}
\label{sec: results}

We explore different methods to visualize how one-pixel attacks and observational noise from exposure time functioned to perturb a base galaxy image in latent space. We use a 150-image sub-sample of our test set of images to perform the one-pixel attack and produce perturbed images. We successfully flip 136 images for both the model trained without and trained with domain adaptation.

\textbf{Church window plots} - these plots derive their name from a colorful representation of latent space around a given input example that resembles panes of stained glass [11, 28, 8]. These colors map onto the potential classes available for classification. To create each plot, the unperturbed base image is situated at the origin. Then, the latent space embedding of the unperturbed base image is subtracted from the embeddings of both the noisy and one-pixel attacked image. These subtractions yield higher dimensional representations of the adversarial and noisy perturbation vectors in latent space (we orient the plane such that the $x$ axis lies along the adversarial direction and the $y$ direction lies along the noisy direction)\footnote{We deviate slightly from the traditional church window plot first introduced by [28], where authors made plots by orienting axes such that the $x$-direction aligned with the adversarial direction and the $y$-direction was simply calculated to be orthonormal to this direction.}. These vectors are then discretized into small steps and we iterate over all possible combinations of both the adversarial and noisy direction perturbations that are added to the embedding of the base image. These new embeddings are then classified by the trained model in order to populate the pixels in the church window plot. Note however, that we plot relative distances between the base image and the noisy and one-pixel attacked image.  

\textbf{Isomaps} - visualizing the data in the multi-dimensional latent space is particularly important and informative for our studies, as it enables the viewing of how specific data points move due to the one-pixel or noisy attack. Here we will use isomaps [26], which seek a lower-dimensional embedding which maintains geodesic distances between points. Compared to linear projections like PCA [9], isomaps show distances present in the original multi-dimensional space more realistically.

\begin{figure}
	\includegraphics[width=\columnwidth]{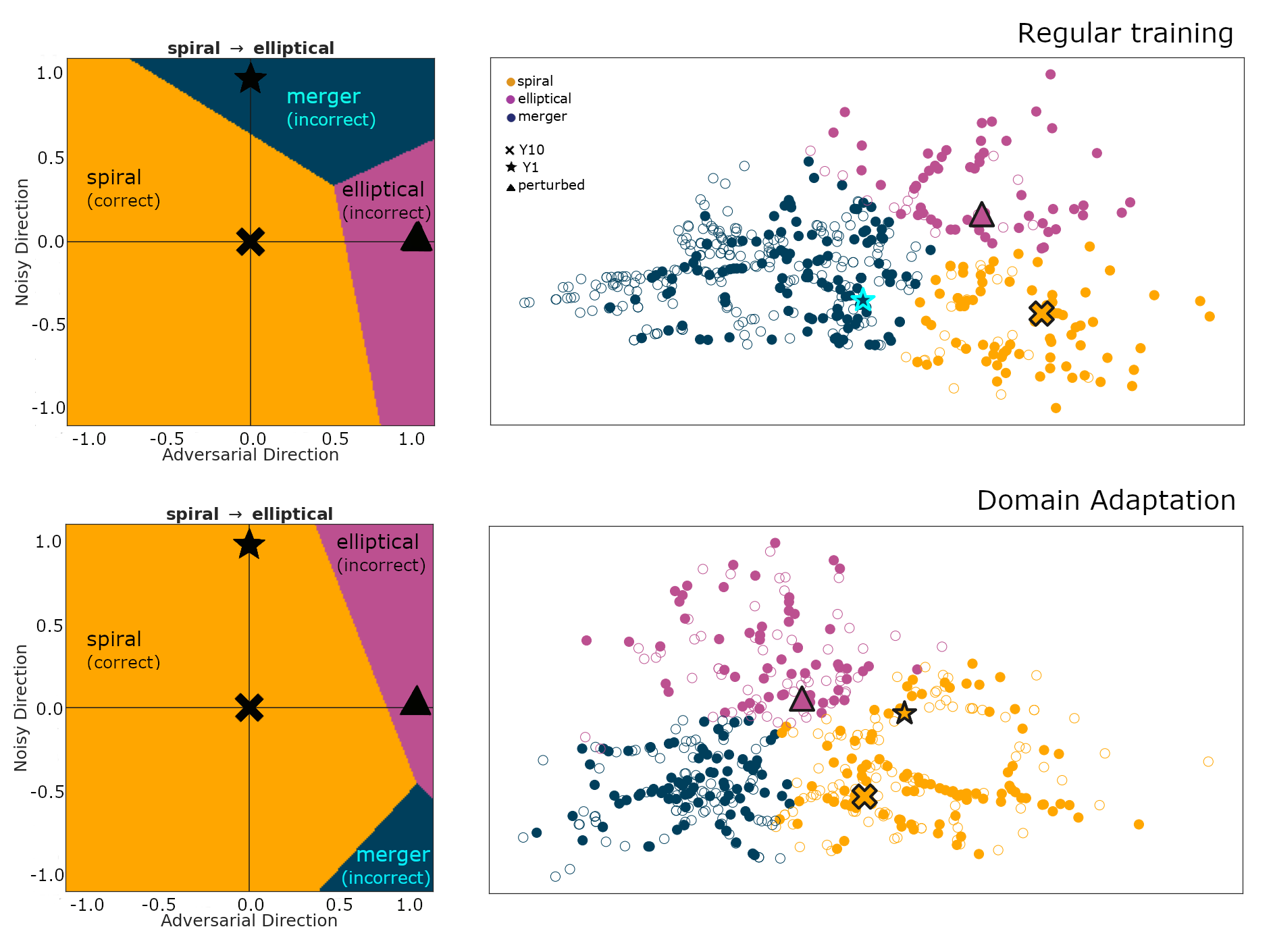}\\
    \caption{Church window plot (left) of example triplet of Y10, Y1 and one-pixel perturbed image (spiral - dark yellow, elliptical - violet, merger - navy), with the 2D isomap (plotted using $250$ randomly selected images from our test set) on the right (Y10 - filled circles, Y1 - empty circles). Each church window plot is labeled with the class of the Y10 image and the class targeted by the one-pixel attack. The Y10 image located at the center of the church window plot is plotted as $\times$ sign in the church window plot and the corresponding isomap, while the noisy Y1 image (top point on the church window plot) is plotted as a star in both plots. Finally, one-pixel attacked image (right point on the church window plot) is plotted as triangle. Bottom row of images shows the same triplet of images, but for a model trained with domain adaptation.}
\end{figure}

\begin{table}
   \centering
   \noindent\begin{minipage}[b]{0.99\columnwidth}
   \centering
    \caption{Performance metrics for \textit{ResNet18} on Y10 and noisy Y1 test data when training without domain adaptation (first row) and with domain adaptation (second row). The table shows the accuracy and weighted precision, recall, and F1 scores. Domain adaptation increases all performance metrics for both Y10 and Y1 data.
    }
  \centering
  \begin{tabular}{|l | l |c c|}
 \hline        &   Metric   &  Y10  & Y1  \\\hline
\multirow{5}{*}{Regular Training}               &  Accuracy     &   $0.72$       &  $0.43$    \\ 
                                                &  Precision    &   $0.76$       &  $0.61$   \\
                                                &  Recall       &   $0.72$       &  $0.43$   \\
                                                &  F1 score     &    $0.72$      &  $0.36$   \\\hline
\multirow{5}{*}{Domain Adaptation}         &  Accuracy     &   $0.82$       &  $0.66$   \\ 
                                                &  Precision    &   $0.82$       &   $0.67$   \\
                                                &  Recall       &   $0.82$       &   $0.66$     \\
                                                &  F1 score     &   $0.82$       &   $0.67$  \\\hline
\end{tabular}
\end{minipage}
\end{table}

\begin{table*}[!ht]
\caption{Means and standard deviations of Euclidean distances in the latent space of the model (between Y10 and noisy Y1 images, as well as Y10 and one-pixel perturbed images), calculated for 136-image sub-sample of the test set.}
\centering
\begin{tabular}{|l | c c |c c|}\hline
         &  Regular   &  Domain Adapt.   \\\hline
    Y10 -- Y1      &      $20.2\pm 11.9$      &   $29.7\pm 17.4$     \\
        Y10 -- One-Pixel     &      $16.8\pm 5.0$      &   $38.1\pm 15.5$     \\\hline
\end{tabular}
\end{table*}

In the top row of Figure 2 we give one example of Y10, Y1 and one-pixel attacked triplet plotted using church window plot (left) and 2D isomap (right). In the bottom row of Figure 2, we plot the same example, but with domain adaptation included in the training. In the church window plots, the morphology predicted by the network is shown by the background color. The Y10 image we plot in the center of the church window plot ($\times$ symbol) is correctly classified as spiral (dark yellow). When regular training is performed the same image, after the one-pixel attack, moves along the $x$ axis to the right (triangle symbol), and is now incorrectly classified as an elliptical (violet). When the image is subject to a noisy attack, it moves upward along the $y$ direction (star symbol), and is now incorrectly classified as a merger (navy). This is because in the multi dimensional latent space regions of the incorrect class are often located near the correctly classified image that the model was trained on. When domain adaptation is used (bottom row of images), the overall performance of the classifier improves. The model is now able to correctly classify the image as a spiral even in the case of high noise, while it is still fooled by a one-pixel adversarial attack (but more iterations were needed to find the pixel that will successfully flip to the wrong class, and the attacked image is further away from the original Y10 image).

Additionally, the isomap in the top row shows that most of the noisy Y1 images are actually incorrectly classified as mergers. When domain adaptation is used (bottom isomap) Y10 (filled circles) and Y1 (empty circles) data distributions overlap a lot better for all classes, which increases network robustness to noise and leads to increase in performance for both datasets. Increase in the model robustness can be also seen if we look at the classification accuracies. Without domain adaptation model trained on Y10 data reaches accuracy of $72\%$, but for noisy Y1 images accuracy is only $43\%$. When domain adaptation is included accuracy on Y10 is $82\%$, and accuracy on noisy Y1 data increases to a much larger $66\%$. In Table 1 we give accuracies, as well as weighted precision, recall, and F1 scores for models trained without and with domain adaptation. 

To quantify the sensitivity of each model to adversarial and noise perturbations, but also get better qualitative understanding of the model behaviour, we can observe how much images move in the found latent space of the network due to perturbations. We calculate the perturbation movement $d_\mathrm{E}$ as the standard Euclidean distance between the position of the original image and the noisy or perturbed image. We find the distribution of these distances for the 136-image test sub-sample, and in Table 2 report the mean and standard deviation of these movements. When the domain adaptation is used, images that were successfully flipped by one-pixel attack needed to move $\sim2.3$ times further from the original Y10 image, which can help increase robustness and improve model performance. 

In conclusion, we showed that naturally occurring adversarial examples that might arise in complex scientific data pipelines can indeed be a big problem in the sciences. Further development and implementation of domain adaptation and other techniques shows great promise in helping us build and implement more robust and trustworthy models.

\section*{Broader Impact}

Our paper demonstrates how noise and instrument problems deteriorate the performance of deep learning algorithms used in science, and the capability of domain adaptation techniques to improve model robustness to this type of issues. This research will impact the astronomy community, but also the wider scientific community, since drop in model performance due to noise or instrument problems is often present in many scientific applications. This is also relevant when simultaneously working with different data sets (for example astronomical observation from different observing years or working with simulated and real data)-- domain transfer problems.

\begin{ack} 
This manuscript has been supported by Fermi Research Alliance, LLC under Contract No. DE-AC02-07CH11359 with the U.S.\ Department of Energy (DOE), Office of Science, Office of High Energy Physics. This research has been partially supported by the High Velocity Artificial Intelligence grant as part of the DOE High Energy Physics Computational HEP program. 
This research has been partially supported by the DOE Office of Science, Office of Advanced Scientific Computing Research, applied mathematics and SciDAC programs under Contract No.\ DE-AC02-06CH11357. 
This research used resources of the Argonne Leadership Computing Facility at Argonne National Laboratory, which is a user facility supported by the DOE Office of Science.

The authors of this paper have committed themselves to performing this work in an equitable, inclusive, and just environment, and we hold ourselves accountable, believing that the best science is contingent on a good research environment.
We acknowledge the Deep Skies Lab as a community of multi-domain experts and collaborators who have facilitated an environment of open discussion, idea-generation, and collaboration. This community was important for the development of this project.

We are very thankful to N.\ Ford for very useful discussion regarding church window plots. 
\end{ack}

\medskip
\small


\section*{References}

[1] Pin-Yu Chen, Huan Zhang, Yash Sharma, Jinfeng Yi, and Cho-Jui Hsieh. ZOO: Zeroth Order Optimizationbased Black-box Attacks to Deep Neural Networks without Training Substitute Models. arXiv e-prints, page arXiv:1708.03999, August 2017.

[2] Aleksandra \'Ciprijanovi\'c, Diana Kafkes, Kathryn Downey, Sudney Jenkins, Gabriel N. Perdue, Sandeep Madireddy, Travis Johnston, Gregory F. Snyder,and Brian Nord. DeepMerge - II. Building robust deep learning algorithms for merging galaxy identificationacross domains.Monthly Notices of the Royal Astronomical Society, 506(1):677–691, September 2021.

[3] Aleksandra \'Ciprijanovi\'c, Diana Kafkes, Sudney Jenkins, Kathryn Downey, Gabriel N. Perdue, Sandeep Madireddy, Travis Johnston and Brian Nord. Domain adaptation techniques for improved cross-domain study of galaxy mergers. arXiv e-prints, pagearXiv:2011.03591, November 2020

[4] Gabriela Csurka. A Comprehensive Survey on Domain Adaptation for Visual Applications, pages 1–35. Springer International Publishing, Cham, 2017.

[5] Swagatam Das and Ponnuthurai Nagaratnam Suganthan. Differential evolution: A survey of the state-of-the-art.IEEE Transactions on Evolutionary Computation, 15(1):4–31, 2011.

[6] Samuel Dodge and Lina Karam. Understanding How Image Quality Affects Deep Neural Networks.arXive-prints, page arXiv:1604.04004, April 2016.

[7] Samuel Dodge and Lina Karam.  A Study and Comparison of Human and Deep Learning RecognitionPerformance Under Visual Distortions.arXiv e-prints, page arXiv:1705.02498, May 2017.

[8] Nic  Ford,  Justin  Gilmer,  Nicolas  Carlini,  and  Dogus  Cubuk.   Adversarial  Examples  Are  a  NaturalConsequence of Test Error in Noise.arXiv e-prints, page arXiv:1901.10513, January 2019. 

[9] Karl Pearson F.R.S.  Liii. on lines and planes of closest fit to systems of points in space.The London,Edinburgh, and Dublin Philosophical Magazine and Journal of Science, 2(11):559–572, 1901.

[10] Milind S. Gide, Samuel F. Dodge, and Lina J. Karam. The Effect of Distortions on the Prediction of VisualAttention. arXiv e-prints, page arXiv:1604.03882, April 2016.

[11] Ian J. Goodfellow, Jonathon Shlens, and Christian Szegedy.  Explaining and Harnessing AdversarialExamples.arXiv e-prints, page arXiv:1412.6572, December 2014.

[12] A. Gretton, Karsten M. Borgwardt, Malte J. Rasch, Bernhard Schölkopf, and Alexander Smola. A kerneltwo-sample test.Journal of Machine Learning Research, 13(25):723–773, 2012.

[13] Arthur Gretton, Dino Sejdinovic, Heiko Strathmann, Sivaraman Balakrishnan, Massimiliano Pontil, KenjiFukumizu, and Bharath K. Sriperumbudur.  Optimal kernel choice for large-scale two-sample tests.  InF. Pereira, C. J. C. Burges, L. Bottou, and K. Q. Weinberger, editors, Advances in Neural InformationProcessing Systems 25, pages 1205–1213. Curran Associates, Inc., 2012.

[14] Kaiming He, Xiangyu Zhang, Shaoqing Ren, and Jian Sun. Deep residual learning for image recognition.In2016 IEEE Conference on Computer Vision and Pattern Recognition (CVPR), pages 770–778, 2016.

[15] \v Zeljko Ivezi\'c, Steven M. Kahn, J. Anthony Tyson, Bob Abel, Emily Acosta, Robyn Allsman, DavidAlonso, Yusra AlSayyad, and et al.  LSST: From Science Drivers to Reference Design and AnticipatedData Products. ApJ, 873(2):111, March 2019.

[16] Diederik P. Kingma and Jimmy Ba. Adam: A method for stochastic optimization.CoRR, abs/1412.6980,2015.

[17] Jennifer M. Lotz, Joel Primack, and Piero Madau. A New Nonparametric Approach to Galaxy Morpholog-ical Classification.The Astronomical Journal, 128(1):163–182, July 2004.

[18] Dylan Nelson, Volker Springel, Annalisa Pillepich, Vicente Rodriguez-Gomez, Paul Torrey, Shy Genel,Mark Vogelsberger, Ruediger Pakmor, Federico Marinacci, Rainer Weinberger, Luke Kelley, Mark Lovell,Benedikt Diemer, and Lars Hernquist. The IllustrisTNG simulations: public data release.ComputationalAstrophysics and Cosmology, 6(1):2, May 2019.

[19] Arjun Nitin Bhagoji, Warren He, Bo Li, and Dawn Song. Exploring the Space of Black-box Attacks onDeep Neural Networks.arXiv e-prints, page arXiv:1712.09491, December 2017.

[20] Vicente Rodriguez-Gomez,  Gregory F. Snyder,  Jennifer M. Lotz,  Dylan Nelson,  Annalisa Pillepich,Volker Springel, Shy Genel, Rainer Weinberger, Sandro Tacchella, Rüdiger Pakmor, Paul Torrey, FedericoMarinacci, Mark Vogelsberger, Lars Hernquist, and David A. Thilker. The optical morphologies of galaxiesin the IllustrisTNG simulation: a comparison to Pan-STARRS observations.Monthly Notices of the RoyalAstronomical Society, 483(3):4140–4159, March 2019.

[21] B. T. P. Rowe, M. Jarvis, R. Mandelbaum, G. M. Bernstein, J. Bosch, M. Simet, J. E. Meyers, T. Kacprzak,R. Nakajima, J. Zuntz, H. Miyatake, J. P. Dietrich, R. Armstrong, P. Melchior, and M. S. S. Gill. GALSIM:The modular galaxy image simulation toolkit. Astronomy and Computing, 10:121–150, April 2015.

[22] A. Smola, A. Gretton, L. Song, and B. Schölkopf.   A hilbert space embedding for distributions.   InAlgorithmic Learning Theory, Lecture Notes in Computer Science 4754, pages 13–31, Berlin, Germany,October 2007. Max-Planck-Gesellschaft, Springer.

[23] R. Storn and K. Price. Differential evolution–a simple and efficient heuristic for global optimization overcontinuous spaces.Journal of global optimization, 11(4):341–359, 1997.

[24] Jiawei Su, Danilo Vasconcellos Vargas, and Kouichi Sakurai.  One pixel attack for fooling deep neuralnetworks. IEEE Transactions on Evolutionary Computation, 23(5):828–841, 2019.

[25] Christian Szegedy, Wojciech Zaremba, Ilya Sutskever, Joan Bruna, Dumitru Erhan, Ian Goodfellow, andRob Fergus. Intriguing properties of neural networks. arXiv e-prints, page arXiv:1312.6199, December2013.

[26] Joshua B. Tenenbaum, Vin de Silva, and John C. Langford. A Global Geometric Framework for NonlinearDimensionality Reduction.Science, 290(5500):2319–2323, December 2000.

[27] Mei Wang and Weihong Deng.  Deep visual domain adaptation: A survey.Neurocomputing, 312:135 –153, 2018.

[28] David Warde-Farley and Ian Goodfellow. Adversarial Perturbations of Deep Neural Networks, pages311–342. MIT Press, 2017.

[29] Garrett Wilson and Diane J. Cook. A survey of unsupervised deep domain adaptation.ACM Trans. Intell.Syst. Technol., 11(5), July 2020.

[30] Xiaoyong Yuan, Pan He, Qile Zhu, and Xiaolin Li. Adversarial examples: Attacks and defenses for deeplearning.IEEE Transactions on Neural Networks and Learning Systems, 30(9):2805–2824, 2019
\clearpage

\end{document}